\documentclass[runningheads]{llncs}
\usepackage{graphicx}
\usepackage{amsmath,amssymb,latexsym}
\usepackage{pifont}
\usepackage{xspace}

\usepackage{import}
\usepackage{graphicx}
\usepackage{multirow}
\usepackage{booktabs}
\usepackage{textgreek}
\usepackage[dvipsnames]{xcolor}
\usepackage[neverdecrease]{paralist}
\usepackage{caption}
\usepackage{subcaption}

\usepackage{rotating}

\usepackage{balance}

\usepackage{subcaption}
\newcommand{\const}[1]{\ensuremath{\mathsf{#1}\xspace}}
  
\usepackage{versions}
\excludeversion{RedundantContent}
\excludeversion{notusec}

\begin{document}
\title{\emph{Dis}-Empowerment Online: \\
An Investigation of Privacy-Sharing Perceptions \& Method Preferences}
\titlerunning{Privacy \& Sharing Empowerment Perceptions}
%
%
\author{Kovila P.L. Coopamootoo \thanks{Kovila P.L. Coopamootoo, Dis-Empowerment Online - An Investigation of Privacy \& Sharing Perceptions \& Method Preferences:  Proceedings of AsiaUSEC'20, Financial Cryptography and Data Security 2020 (FC). February 14, 2020 Kota Kinabalu, Sabah, Malaysia Springer, 2020} }

\institute{Newcastle University, UK \\
\email{kovila.coopamootoo@newcastle.ac.uk}}

\maketitle              
%

\begin{abstract}
While it is often claimed that users are empowered via online technologies, there is also a general feeling of privacy dis-empowerment. We investigate the perception of privacy and sharing empowerment online, as well as the use of privacy technologies, 
via a cross-national online study with N=907 participants. 
We find that perception of privacy empowerment differs from that of sharing across dimensions of meaningfulness, competence and choice.
We find similarities and differences in privacy method preference between the US, UK and Germany. 
We also find that non-technology methods of privacy protection are among the most preferred methods, while more advanced and standalone privacy technologies are least preferred..
By mapping the perception of privacy dis-empowerment into patterns of privacy behavior online,
and clarifying the similarities and distinctions in privacy technology use, this paper provides an important foundation for future research and the design of privacy technologies. The findings may be used across disciplines to develop more user-centric privacy technologies, that support and enable the user.

\end{abstract}
%
%
%

\section{Introduction}
\begin{notusec}
The internet is often seen as an empowering environment for consumers impacting personal, interpersonal, group and citizen-wide dynamics~\cite{amichai2008empowerment,fuglsang2005and}, and enabling consumer influence on product design, choice and decisions~\cite{fuller2009consumer}, as well as co-creation~\cite{wathieu2002consumer}.
However, the indiscriminate amount of information collected for this purpose is also seen to come with privacy-, identity- and empowerment-related issues~\cite{o2008lifelogging}. 

Indeed, today's data-intensive web is characterized with mass sharing, collection, aggregation, mining and selling of individuals' data. 
It enables the provisioning of customised services but unfortunately also engenders targeted advertising~\cite{chandramouli2013real}, digital discrimination~\cite{edelman2014digital}, privacy invasive algorithmic computations~\cite{forbes2019}, and a general fuzzyness about privacy rights online.
Various high profile cases involving mass unauthorised transfer and use of sensitive data have been reported in the recent years, including influencing voters via social media~\cite{garrett2019social}, top health websites' are sharing sensitive data with advertisers, including Google and Facebook~\cite{murgia2019how} as investigated by the Financial Times, and period apps' (such as Maya and MIA) extensive sharing of sensitive personal data pertaining to women's health with third parties, including Facebook as found by Privacy International~\cite{privacyint2019nobodys}.
\end{notusec}

Although the internet is often seen as an empowering environment for consumers, the indiscriminate amount of information collected in today's data-intensive web, characterized with mass sharing, collection, aggregation, mining and selling of individuals' data, 
is also seen to come with privacy-, identity- and empowerment-related issues~\cite{o2008lifelogging}. 

Internet users often express discomfort with the data collection that enables personalization, and a large portion takes some kind of action such as clearing cookies and browsing history~\cite{rainie2013anonymity}.
However, the methods employed by individuals may not be enough to protect one's privacy, because, as example, a particular web browser on a specific machine comprises a unique fingerprint that can be traced by web servers across the web, and this information in conveyed through headers that are automatically exchanged by every web browser and web server behind the scenes~\cite{nikiforakis2013cookieless}.

In general, privacy experts perceive an overall sense of privacy dis-empowerment online~\cite{coopamootoo2018towards}.
The perception of privacy \emph{dis}-empowerment has mainly been attributed to business models and the social web that favour sharing and data analytics, to privacy of personal content~\cite{mansell2014empowerment,pierson2012online}.
Other reasons include human challenges to the adoption of privacy technologies, 
and human-computer mismatches. 
We posit that privacy dis-empowerment is evidenced in the failure to use privacy technologies~\cite{abu2017obstacles,harborth2018examining}.

\begin{notusec}
When individuals share online, they are willing to take a number of steps towards social or interpersonal privacy for data shared as part of social interactions. 
These include the use of privacy settings for boundary regulation, as well as coping strategies to reduce emotional distress from privacy loss~\cite{wisniewski2012fighting}, such as blocking others, creating multiple profiles, censoring themselves, untagging and offline negotiations~\cite{stutzman2012boundary}.

Yet, individuals may not perceive the connection between social and informational privacy, where strategies such as specifying privacy settings or having multiple profiles do not support better control over informational privacy, because the architectures and algorithms that collect data and make inferences are mostly invisible to users~\cite{de2005two}. 
Privacy protection is therefore perceived to be a `losing game' for individuals as more and more data about them is being generated faster and faster from more and more devices~\cite{kerry2019why}. 
\end{notusec}

Use of privacy technologies is thought to be moderated by user perception of technology. In particular, perceived usefulness and effectiveness do not match the technology's offering, and users exhibit poor trust in the technology~\cite{abu2017obstacles,benenson2015user,harborth2018examining}, and in-correct mental models~\cite{abu2017obstacles}.
However, individuals are likely impacted by their own self-perception, in addition to their perception of the technology.
As a result, they likely engage with some privacy technologies more than others, employ privacy technologies in a certain way, or develop non-technology methods of protection.


\emph{\textbf{Contributions:}}
In this paper, we seek to better understand how the perception of privacy (dis)-empowerment is mapped out into
patterns of privacy behavior online. 
We employ a quantitative method as we investigate
how individuals protect their privacy from others - whether individual others or organisations - 
in particular what privacy methods they use. 
We investigate the link between perception of dis-empowerment and behavior across 40+ privacy methods elicited from users themselves. 
The paper makes the following contributions:
(1) We provide a cross-national report of users' perception of empowerment. 
(2) We find that individuals use $22$ privacy methods on average, where $40$ to $50\%$ of the $10$ topmost preferred methods are non-technology methods that are 
reported to be used by $71\%$ to $85\%$ of the surveyed participants.
\begin{notusec}
(4) We find that the use of an additional privacy method is only likely with an increase in perceived competency of $18.5$-points and perceived meaningfulness of $17$-points on a visual analogue scale. Men use approximately $1.5$ more privacy method than women.
\end{notusec}
(3) We identify similarities and differences in privacy and sharing method preferences between the three countries.


This paper therefore provides valuable insights into individuals' methods of protecting their privacy online, that includes both non-technology methods and the use of privacy technologies.
This helps to ground the perceptions of privacy dis-empowerment into behavior patterns. 
The paper also helps to identify privacy technologies that appear to be more accessible to users. 



\emph{\textbf{Outline:}}
In the rest of the paper, 
provide the aim and method of our study, followed by the results and a discussion, and conclusion. 

\section{Aim}
Our research aim is to compare privacy and sharing empowerment perceptions and to map perceptions of privacy \emph{dis}-empowerment onto usage of privacy and sharing methods. We do so via the research questions below.

\begin{notusec}
\begin{figure}[h]
\centering
\includegraphics[keepaspectratio,width=.95\columnwidth]{./figures/concept}
\caption{Research Summary.}
\label{fig:concept}
\end{figure}
\end{notusec}

\subsection{Privacy vs Sharing Empowerment}
Thomas \& Velthouse~\cite{thomas1990cognitive} defined \emph{Psychological Empowerment} as increased intrinsic task motivation and proposed a theoretical model with four perceptions or cognitions, namely \emph{perception of impact}, \emph{competence}, \emph{meaningfulness}, and \emph{choice}~\cite{thomas1990cognitive}. 
The model captures individuals' interpretive processes via which they assess the actions they engage in. 
Compared to other psychological empowerment models, Thomas \& Velthouse's model focuses on intrinsic motivation and involves positively valued experiences that individuals derive directly from a task, and impact behavior.


With the power imbalance between online users and others (including more able other individuals perceived as threatening and organisations), individuals likely perceive privacy and sharing empowerment differently online. 
We investigate as \textbf{RQ1}, 
``How do individuals' perception of privacy and sharing empowerment differ?" via the hypotheses: \\
\const{H_1,_0}:There is no difference in individuals' perception of privacy and sharing empowerment.\\
\const{H_1,_1}:There is a significant difference in individuals' perception of privacy and sharing empowerment.

\subsection{Privacy \& Sharing Methods, Similarties \& Differences}
We investigate as \textbf{RQ2},
``What methods are mostly used to protect one's privacy and to share information online?"
and \textbf{RQ3}, 
``How similar are individuals' [privacy/sharing] methods usage and preference?
What patterns of use emerge? Are there similarities or differences between countries?" 

\section{Method}
We conduct two survey studies online via an evidence-based method~\cite{coopamootoo2016evidence,coopamootoo2017ifip}. 
The first study is mainly aimed at identifying a preferred list of privacy methods.
The second and main study employs the compiled list of methods to query a representative sample of participants about their use of the range of privacy methods identified.

The studies have a within subject design, where participants answered both the privacy and sharing empowerment questions. We compared privacy and sharing empowerment for each participant. However, we compared preferred privacy and sharing methods between countries, thereby including a between-subject analysis. 
We randomly assigned participants to answer either the privacy or sharing empowerment questions first.

\begin{RedundantContent}
We conduct a first study with $N=180$ participants, querying participants about empowerment perception and their preferred list of privacy and sharing methods. 
We compile lists of preferred privacy and sharing methods. 
Using these lists, we design and run a second study with a representative sample with $N=606$ participants. We ask participants to select their preferences from the lists provided. 
\end{RedundantContent}

\subsection{Participants}
\begin{notusec}
With their advanced digital economies, Europe and the USA may be considered drivers of protection technology around the globe.
However, Europe versus the US differ in privacy regulation~\cite{pwc2016data} and individuals' privacy protection culture.
\end{notusec}

\subsubsection{Recruitment}
For the first study, we sampled  $N=180$ participants, comprising $N=58$ US participants, $N=62$ UK participants and $N=60$ German (DE) participants. The US sample was recruited from population of Amazon Mechanical Turk workers, while the UK and DE sample were from Prolific Academic. 
The data quality of Prolific Academic is comparable to Amazon Mechanical Turk's, with good reproducibility~\cite{peer2017beyond}.

For the second study, we recruited an $N=907$ sample from the US, UK and DE via Prolific Academic. The sample was representative of age, gender and ethnicity demographics of the UK and US countries, as provided by Prolific Academic. 
For the DE sample, we did not achieve a representative sample in terms of gender and age.
While we use that sample to investigate our research questions, we foresee extending to representative samples of other countries in the future.

The studies lasted between $10$ to $20$ minutes. Participants were compensated at a rate of \pounds$7.5$ per hour, slightly above the minimum rate of \pounds$5$ per hour suggested by Prolific Academic.

\subsubsection{Demographics}
Table~\ref{tab:demographics} provides a summary of the demographic details for the two studies, with sample size $N$, mean age, gender, education level and ethnicity.
$5\%$ of the German sample had an education level lower than high school for the first study and $1\%$ for the second study.
For the second study, $6$ UK participants reported to have a PhD, $4$ for the US and $9$ for DE.

\begin{table*}[h]
\vspace{-.5cm}
\centering
\caption{Participant Characteristics}
\label{tab:demographics}
\footnotesize
\resizebox{\textwidth}{!}{
\begin{tabular}{lcrcrrcrrrrlrrrrr}
\toprule
&\textbf{Country}& \textbf{$N$} & \textbf{Mean Age} & \multicolumn{2}{c}{\textbf{Gender}} &&\multicolumn{4}{c}{\textbf{\%Education Level}} &&\multicolumn{5}{c}{\textbf{\% Ethnicity}} \\
\cline{5-6}
\cline{8-11}
\cline{13-17}
&&&& \#Female & \#Male &&HighSchool&College&Undergrad&Masters/PhD && White & Black & Asian & Mixed & Other\\
\midrule
\multirow{3}{*}{\textbf{First Study}}&US & $58$ & $35.53$ & $29$ & $29$ && 24.1&31.0&36.2&8.6 && 82.8 & 5.2 & 5.1&5.2&1.7\\
&UK & $62$ & $30.65$ & $43$ & $19$ && 22.6&19.4&41.9&16.1&& 88.7 &3.2&3.2&4.8&-\\
&DE & $60$ & $30.68$ & $27$ & $33$ && 30.0 & 13.3 & 28.3 & 21.7&&96.7&-&-&3.3&-\\
\midrule 
\multirow{3}{*}{\textbf{Second Study}}& US &303 & 43.72&155&148 &&39.9 &22.1 &20.1 &14.2&&69.3&14.9&8.9&4.3&2.6\\
&UK & 303 & 44.21 &154 &149 &&26.7&17.5&32.0&18.5 && 77.6 &5.3 & 10.9 & 4.3 & 2.0\\
&DE & 301 & 28.91 & 115 & 186 && 31.2 & 15.6 & 28.6 & 23.6&& 93.0&0.7&1.9&3.7&0.7\\
\bottomrule
\end{tabular}
}\\
\vspace{-1cm}
\end{table*}

\subsection{Procedure}
\label{sec:procedure}
The aim of the first study was to identify and compile a list of privacy and sharing methods preference. 
We did so via an open-ended question and across three countries.
The first study consisted of \begin{inparaenum}[(a)]
\item a questionnaire on demographics, 
\item a description of privacy online, and the four psychological empowerment questions, 
\item an open-ended query to list three to five tools most often employed to achieve the purpose of privacy online, 
\item a description of sharing online, and the four psychological empowerment questions,
\item an open-ended query to list three to five tools most often employed to achieve the purpose of sharing online. 
\end{inparaenum}

The second study followed the same format of the first study, except that we changed the open-ended queries of the first study to close-ended privacy and sharing methods questions, for participants to select the methods they mostly use from the whole list provided.
We also shifted to a larger sample for the three countries. 

We defined privacy and sharing for the two studies, thereby focusing participants to a specific meaning.
We developed the definition of [privacy/sharing] online with inputs from Coopamootoo \& Gro{\ss}'s findings of the cognitive content of individuals' [privacy/sharing] attitude~\cite{coopamootoo2017whyprivacy}. 
In particular, privacy attitude has contents of `others as individuals or organisations who pose a threat, 
while sharing attitude includes `others as connections including friends, family'. 

We defined privacy online as 
\emph{``to control access to information that are sensitive or personal, to be informed of other individual and business practices such as collection, processing and use of personal information disclosed, and to have the choice on disclosure and how one's information is dealt with."}
We defined Sharing online as 
\emph{``to create content and share with other web users (such as sharing one's opinion or expertise) and also to share personal information or life events with close connections, friends and family."}
\begin{notusec}
We provide a summary of the procedure in Figure~\ref{fig:structure}.
\begin{figure*}
\vspace{-.5cm}
\centering
\includegraphics[scale=0.7]{./figures/experiment_structure_v3}
\caption{Experiment design including user study, expert evaluation and subsequent analyses.}
\label{fig:structure}
\vspace{-1.1cm}
\end{figure*}
\end{notusec}

\subsection{Measurement Apparatus}
\subsubsection{Perception of Psychological Empowerment}
Measures of psychological empowerment have mainly been employed within management and social science research~\cite{spreitzer1995psychological,menon1999psychological}.
In particular, Spreitzer proposed a four-factor scale based on Thomas \& Velthouse's conceptualization~\cite{thomas1990cognitive}. The scale has been widely applied in the context of organizational management~\cite{spreitzer1995psychological}. 
It has also been evaluated for construct validity~\cite{kraimer1999psychological}.
In addition, Spreitzer's formulation was observed as seminal to research on psychological empowerment~\cite{seibert2011antecedents}. 
However, so far, sparse application appear in relation to technology, such as Van Dyke et al.'s measure of consumer privacy empowerment in E-Commerce~\cite{van2007effect}.

The Psychological Empowerment Scale consists of $12$-items focused on the four dimensions of empowerment defined by Thomas \& Velthouse~\cite{thomas1990cognitive}, in particular areas of \begin{inparaenum}[(1)]
\item meaning,
\item competence, 
\item self-determination/choice, and
\item impact.
\end{inparaenum}
Whereas Van Dyke et al. apply these four dimensions to the notice, choice and access concepts to then develop four perceived privacy empowerment items~\cite{van2007effect}, we directly adapted Spreitzer's scale~\cite{spreitzer1995psychological} for online [privacy/sharing] activities.
We used the four cognitions of the model to create a task assessment questionnaire directed towards the purpose of either privacy or sharing online. 

We set the [privacy/sharing] questions as follows: \\
\emph{`Purpose' refers to that of achieving [privacy/sharing] online as detailed above.\\
`Actions' are those that one would take with the aim to accomplish that purpose, that is [privacy/sharing] online.\\
Please provide your responses on the scale from 1 to 100.}
\begin{inparaenum}[(1)]\\
\item \emph{How do you perceive the impact of the actions you have taken online in the past to accomplish the purpose detailed above?}\\
\item \emph{How do you perceive your skills to successfully achieve the purpose detailed above?}\\
\item \emph{To what extent is the purpose detailed above meaningful to you?}\\
\item \emph{How do you perceive your choice to determine what actions to take to successfully accomplish the purpose detailed above?}
\end{inparaenum}

We used a Visual Analogue Scale (VAS)~\cite{wewers1990critical}
with boundaries from $1$ to $100$.
The $1$ minimum value was set to `no impact at all', `not skilled at all', `not meaningful at all' or `I have no choice at all', pertainig to the four questions above.
The $100$ maximum value was set to very `big impact', `very skillful', `very meaningful'' or `I have lots of choices. 

Compared to Likert-type scales which have coarse-grained discrete measurement data produced by only three to seven categories, the line continuum of a VAS enables the rater to make more fine-grained responses~\cite{chimi2009likert}.
This aspect of VAS helps to avoid the systematic bias of values resulting from scale coarseness~\cite{aguinis2009scale}
and facilitates collection of measurement data with higher variability, which theoretically enhances their reliability~\cite{cook2001score}. 

\subsubsection{Privacy \& Sharing Behavior}
We queried participants on the individual privacy and sharing methods they most often use, eliciting participants' own methods via open-ended question in the first study and requesting preference report from the compiled list in the second study.
In the second study, we asked participants to rate the list of privacy and sharing methods provided with whether they use them `very often' or `very rarely/not at all'.

\begin{notusec}
In addition, we compute privacy behavior as the total number of different methods participants employ to protect their privacy online.
\end{notusec}

\section{Results}

\subsection{Empowerment Perception}
We investigate \textsf{RQ1} with respect to the US, UK and DE samples in the second study, ``How do individuals' perception of privacy versus sharing empowerment differ?"
We conduct a paired-samples $t$-test for privacy and sharing for each of the four cognitions for the three countries.
We summarize the differences in perception of privacy and sharing empowerment cognitions in Table~\ref{tab:psych_emp} below.

\begin{table*}[h] 
\centering
\caption{Task Assessment Differences between Privacy Activities \& Sharing Activities} 
\label{tab:psych_emp}
\footnotesize
\resizebox{\textwidth}{!}{\begin{tabular}{llllllrr@{}lrrrrrr}
\toprule
\multirow{2}{*}{Assessment Component} & \multicolumn{2}{c}{\const{Privacy}} & &\multicolumn{2}{c}{\const{Sharing}} & \multirow{2}{*}{$t(df)$} & \multirow{2}{*}{$p$} & &\multicolumn{2}{c}{Difference} & & \multicolumn{2}{c}{$95\%$ CI} & \multirow{2}{*}{}\\
\cline{2-3}
\cline{5-6}
\cline{10-11}
\cline{13-14}
& $M$ & $SD$ & & $M$ & $SD$ &&&& $\Delta{}M$ & $SE$ && LL & UL \\
\midrule
\textbf{United States} &  & &  &&  &$t(302)$& &  & &&&& \\ 
&  & &  &&  && &  & &&&&\\
 Meaningfulness & $76.36$ & $21.739$ && $63.94$ & $30.823$ & $8.489$ & $<.000$ &*** & $15.419$ & $1.816$ & & $11.845$ & $18.993$ &  \\
Competency &$58.81$&$23.508$&&$66.05$&$24.380$&$-5.087$&$<.000$&***&$-7.238$&$1.423$&&$-10.037$&$-4.438$&\\ 
Choice &$61.78$&$22.397$&&$72.49$&$22.532$&$-7.331$&$<.000$&***&$-10.706$&$1.460$&&$-13.580$&$-7.832$&\\
Impact &$58.16$&$22.489$&&$58.65$&$25.862$&$-.287$&$.774$&&$-.498$&$1.734$&&$-3.911$&$2.914$& \\
\midrule
\textbf{United Kingdom} &  & &  &&  &$t(302)$& &  & &&&& \\ 
&  & &  &&  && &  & &&&&\\
Meaningfulness &$70.06$&$24.786$&&$59.84$&$26.875$&$5.862$&$<.000$&***&$10.211$&$1.742$&&$6.783$&$13.639$&\\
Competence&$56.87$&$22.714$&&$62.20$&$23.084$&$-4.013$&$<.000$&***&$-5.330$&$1.328$&&$-7.944$&$-2.716$&\\
Choice& $59.10$ & $21.562$ && $66.81$ & $21.907$ & $-5.747$ & $<.000$ &***& $-7.716$ & $1.343$ && $-10.358$ & $-5.047$ &\\
Impact &$54.79$&$21.885$&&$57.42$&$24.265$&$-1.604$&$.110$&&$-2.637$&$1.644$&&$-5.872$&$.598$&\\
\midrule
\textbf{Germany} &  & &  &&  &$t(300)$& &  & &&&& \\ 
&  & &  &&  && &  & &&&&\\
Meaningfulness &$69.63$&$22.998$&&$48.59$&$29.984$&$9.947$&$<.000$&***&$21.040$&$2.115$&&$16.877$&$25.202$&\\
Competence & $58.73$&$23.280$&&$62.85$&$24.678$&$-2.662$&$.008$&**&$-4.123$&$1.549$&&$-7.171$&$-1.075$& \\
Choice & $55.57$&$19.782$&&$68.94$&$23.202$&$-8.870$&$<.000$&***&$-13.365$ &$1.507$&&$-16.331$&$-10.400$\\
Impact &$53.26$&$21.793$&&$49.03$&$24.558$&$2.550$&$.011$&*&$4.223$&$1.656$&&$.964$&$7.481$ \\

\bottomrule
\end{tabular}}\\
\emph{CI} refers to the Confidence Interval, LL to the Lower Limit, UL to the Upper Limit.\\
\end{table*}

\subsection{Privacy \& Sharing Methods}
We provide the full list of privacy methods compiled in the first study in Table~\ref{tab:all_privacy_methods}, with the $N=180$ sample.
This list of $43$ privacy methods was then used to query participants in the second study.
We loosely categorise the privacy methods into four possible protection categories, namely \begin{inparaenum}[(1)]
\item anonymity (ANO), 
\item browsing history and tracking prevention (BHP), 
\item communication privacy \& filtering (COP), and
\item preventing leaking and stealing of data (PLS).
\end{inparaenum}

We also compile participants' responses of $3$ to $5$ most used sharing methods in the first study.
We end up with $39$ sharing methods coded across the three countries.
We categorize the sharing methods across five themes, as shown in Table~\ref{tab:all_sharing_methods}.
The `Community' theme includes social networks or community sharing. With respect to Facebook, some participants referred to Facebook in general, while others specified updates or photos.
The `Messaging' theme includes email and instant messaging methods, referring to a particular tool or instant messaging in general.
The other sharing themes are `Photos', `File-Sharing' and `Streaming'.

The rest of the results section pertains to the second and main study.

\begin{table*} [h]
\centering
\footnotesize
\caption{Privacy Methods Categorised by Design Type and Privacy Protection.}
\label{tab:all_privacy_methods}
\resizebox{\textwidth}{!}{
\begin{tabular}{llll} 
\toprule
\textbf{Privacy Protection} & \textbf{Built-in} & \textbf{Standalone} & \textbf{User-Defined}\\

\midrule
\multirow{10}{*}{\textbf{Anonymity}} &Encryption&Erasery&Not Store Info\\
&Clear/Delete info/history&TOR&Anonymous profile names\\
&Pseudonyms/Onion&Proxy&NotGivePI / LimitSharing / MinimalInfo\\
&&IPHider&Several/Bogus / LimitedUse Emails\\
&&Virtual machine&Fake Info\\
&&&Limit Use of SNS Accounts\\ 
&&&SwitchOffCamera/Devices/PortableHD\\
&&&No Access Acc In Public Place/Networks\\  
&&&Not use FB\\
&&&Not Engaging Online/Careful/Not Signing Up\\

\midrule
&Private Browsing/incognito&DuckDuckGo&\\
\textbf{Browsing History \&} &Anti-tracking addon&Ghostery&\\
\textbf{Tracking Prevention} &No location tracking &NoScript&\\
&Clear/Limit cookies&&\\

\midrule
\textbf{Communication \&} &Adblock&Firewall&\\
\textbf{Filtering} &HTTPS&VPN&\\

\midrule
&Privacy settings &Password manager&Not save or reuse password\\
\textbf{Prevent Leaking \&}&Opt out &Paypal&Read terms of service\\ 
\textbf{Stealing of Data}  &Private profiles &Anti-spyware& Request data collected, GDPR\\ 
&&Anti-malware& no newsletter, think twice\\
& &Kapersky& Website care/No suspicious sites\\

\bottomrule
\end{tabular}
}
\end{table*}

\begin{table*} 
\centering
\footnotesize
\caption{Sharing Methods Categorised by Theme.}
\label{tab:all_sharing_methods}
\begin{tabular}{lllll} 
\toprule
\textbf{Community} & \textbf{Photos} & \textbf{Messaging} & \textbf{File-Sharing} & \textbf{Streaming}\\
\midrule
Discord & Facebook photos &Email & box.com&Twitch\\
Facebook & Flickr & Facebook messenger & cloud&Vimeo\\
Facebook updates &Google photos & Instant messaging & dropbox&YouTube\\
Forums &iCloud photos & Telegram &FTP\\
Google hangouts &Instagram &WhatsApp & Google Drive\\
LinkedIn &Social network photos && Microsoft OneDrive\\
News site comments & Pinterest\\
Personal blog & Photo blog\\
Reddit &Snapchat\\
Skype\\
Slack \\
Social networks\\
Social network updates\\
Teamviewer\\
Tumblr\\
Twitter\\

\bottomrule
\end{tabular}
\end{table*}

We investigate \textsf{RQ2}
``What methods are mostly used to protect one's privacy and to share information online?"
How similar are individuals' [privacy/sharing] methods usage and preference?
What patterns of use emerge? Are there similarities or differences between countries?"


Table~\ref{tab:top10} shows a depiction of the top $10$ privacy methods preferences across the three countries, where we observe
that $4$ of the privacy methods appear in the top $10$ most reported methods in all three countries. These methods are 
\begin{inparaenum}[(1)]
\item privacy settings, 
\item limit sharing, 
\item website care, and 
\item no newsletter.
\end{inparaenum}

In addition, we find $8$ privacy methods similarities in the top $10$ most reported methods for both the UK and US, $6$ methods similarities between the UK and DE, and $5$ methods similarities between the US and DE.

\begin{table*}[h]
\centering
\caption{Top 10 Privacy Methods by Country starting with most frequently mentioned}
\label{tab:top10}
\footnotesize
\resizebox{\textwidth}{!}{
\begin{tabular}{l@{}lll c l@{}lll c l@{}lll}
\toprule
\multicolumn{4}{c}{\textbf{United States}}&&\multicolumn{4}{c}{\textbf{United Kingdom}}&&\multicolumn{4}{c}{\textbf{Germany}}\\
\cline{1-4}
\cline{6-9}
\cline{11-14}
\multicolumn{2}{l}{\textbf{Method}}&\textbf{Design}&\textbf{CAT}&&\multicolumn{2}{l}{\textbf{Method}}&\textbf{Design}&\textbf{CAT}&&\multicolumn{2}{l}{\textbf{Method}}&\textbf{Design}&\textbf{CAT}\\
\midrule
1 &Website care &UD&PLS&&1&Website care& UD & PLS &&1&AdBlock&BI&COP\\
2 & Privacy settings &BI&PLS&&2 & Limit Sharing &UD&ANO && 2& Bogus Emails &UD&ANO\\
3 & Limit Sharing &UD&ANO&&3 & Privacy settings&BI&PLS&&3 & Privacy settings&BI&PLS\\
4 & Research before engaging &UD&ANO&&4 & Clear Info/History &BI&ANO && 4&  Limit Sharing &UD&ANO\\
5 & Anti-Malware &ST&PLS&&5 & Paypal&ST&PLS &&5& No Newsletter &UD&PLS\\
6 & No Newsletter &UD&PLS&&6 & Research before engaging&UD&ANO && 5&Paypal&ST&PLS \\
7 & AdBlock &BI&COP&&7 &No Newsletter &UD&PLS&&5 &Website care& UD & PLS \\
8 & Clear Info/History &BI&ANO&&8 & Firewall &ST&COP&&5& Firewall &ST&COP\\
9 & Clear/Limit Cookies &BI&BHP&&9 & Anti-Malware&ST&PLS &&9&HTTPS&BI&COP\\
10 & Not Access Accts in Public Place &UD&ANO&&10 & Not Access Accts in Public Place &UD &ANO &&10& Pseudonyms &BI&ANO\\
\bottomrule 
\end{tabular}
}\\
\emph{BI}, \emph{ST} \& \emph{UD} refer to design type of built-in, standalone and user-defined respectively.\\
\emph{ANO}, \emph{BHP}, \emph{COP} \& \emph{PLS} refer to privacy protection categories of anonymity, browsing history and tracking prevention, communication privacy \& filtering, and preventing leaking \& stealing of data respectively.
\end{table*}

Table~\ref{tab:top10sharing} shows a depiction of the top $10$ sharing methods preferences across the three countries, where we observe that $5$ of the sharing methods appear in the top $10$ of all three countries, and 
$8$ appear in the top $10$ most reported methods for both the US and the UK.

\begin{table*}[h]
\centering
\caption{Top 10 Sharing Methods by Country starting with most frequently mentioned}
\label{tab:top10sharing}
\footnotesize
\begin{tabular}{ll c ll c ll}
\toprule
\multicolumn{2}{l}{\textbf{United States}}&&\multicolumn{2}{l}{\textbf{United Kingdom}}&&\multicolumn{2}{c}{\textbf{Germany}}\\
\midrule
1& Email && 1& Email && 1 &Email\\
2& Youtube && 2& WhatsApp && 2 & WhatsApp\\
3& Google Drive && 3&Facebook Messenger && 3&YouTube\\
4& Facebook Messenger &&4& YouTube  && 4 & Reddit\\
5& Reddit &&5 & Instant Messaging && 5 & Instagram\\
6& Instant Messaging &&6& Facebook updates \& newsfeed && 6 & Google Drive \\
7& Forums&&7& Google Drive && 7& DropBox \\
8& Instagram &&8& Instagram && 8&Instant Messaging \\
9& Facebook updates \& newsfeed &&9& Twitter &&9& Discord\\
10 & Facebook photos&&10 & Facebook photos && 10 & Forums\\
10& Social network sites (exclu. FB) & \\
10& Twitter & \\

\bottomrule
\end{tabular}\\
\end{table*}


We investigate whether there is a difference in privacy method preference between countries.
On average, participants reported to protect their privacy with $22$ different ways ($m=21.86$, $sd=7.11$). 
DE and US participants reported using $3$ and $2$ more privacy methods on average than UK participants respectively

We compute a Chi Square test on each of the $43$ privacy methods.
We find that for $23$ privacy methods, there is a statistically significant association between privacy method employed and country of residence, after multiple comparisons correction, as detailed in Table~\ref{tab:differences} in the Appendix.
The table shows both the percentage of participants within each country who listed the privacy method, as well as the percentage taken by each country for each listed method. It also shows the privacy protection category of the method. 

In addition, the table provides a measure of association in privacy method preference across country of residence, with effect size Cramer $V$ depicting the magnitude of association between the privacy method and the country, 
where $V<.20$ corresponds to a weak association, $.20<V<.40$ corresponds to a moderate association and $V>.40$ corresponds to a strong association. 


\begin{notusec}
We investigate the second part of \textsf{RQ-T1} with respect to sharing methods, that is, ``Is there a difference in [sharing] method preference between the two countries?"
We compute a Fisher Exact test on each of the $39$ sharing methods.
We find that for $12$ sharing methods, there is a statistically significant association between sharing method employed and country of residence, as detailed in Table~\ref{tab:differencessharing}.
A significantly higher number of participants from the US sample reported using Discord, forums, Google Drive, Google Hangout,  Reddit, Tumblr, Twitch, Vimeo and YouTube compared to the UK sample.
In addition, a higher number of participants from the UK sample reported using Facebook messenger, instant messenger in general and WhatsApp compared to the US sample.
From the Cramer $V$ effect sizes, we find that the differences for Reddit and Twitch refer to medium effects where as that for WhatsApp refers to a strong effect. The rest of the methods with differences show a weak effect.

We therefore reject the null hypothesis \const{H_{T1},_0} that ``There is no difference in [sharing] methods preference between the two countries" for the $12$ sharing methods listed in Table~\ref{tab:differencessharing}.
\end{notusec}

\vspace{-.3cm}
\section{Discussion}
\emph{\textbf{Privacy vs Sharing Empowerment:}}
That participants perceive privacy to be more meaningful than sharing across the three countries, yet perceive lower competency and choice with regards to privacy can be expected given the looming sense of privacy dis-empowerment online users are habituated to.
However by providing statistical evidence from a relatively large representative sample, we demonstrate that privacy dis-empowerment is not just a passing or one-time feeling but is perceived across countries and demographics.
This finding can contribute to explaining the privacy paradox, that although individuals are concerned about their privacy, their observed behavior differ, as they have poor perceptions of competency and choice.

In addition, although the internet is thought to empower individuals, we do not observe a positive difference in perceived sharing impact versus perceived privacy impact. This aspect requires further investigation in eliciting users' understanding of the results of their sharing. Only DE shows a higher perceived impact for privacy.\\

\subsection{Methods Preference \& Behavior}
DE and US participants reported using $3$ and $2$ more privacy methods on average than UK participants respectively, where although there are similarities in that $4$ items are among the top $10$ most used privacy methods in countries, they differ across $23$ methods. 
DE shows a higher use of $19$ methods, a higher portion of which are more technologically advanced PETs rather than simpler builtin PETs. 
This may indicate higher awareness of and skill to use PETs, as well as an outcome of privacy culture and regulation.

Among the similarities, we find that user-defined or non-technology methods \begin{inparaenum}[(1)]
\item of being careful of websites, 
\item to limit sharing, 
\item research before engaging (2 out of 3 countries), 
\item not subscribe to newsletters, and 
\item not access accounts in public places appear in the most used methods in both countries. 
\end{inparaenum} 
For the three countries, these non-technology methods made up $40\%$ to $50\%$ of the top $10$ most preferred privacy methods, while advanced, dedicated and standalone PETs such as Tor, Ghostery or NoScript are among the least used privacy methods.
This demonstrates that users rely more on their own non-technology means to protect themselves than privacy technologies.

Questions can be raised for future research following these usage patterns.
In particular, ``what are reasons for reliance on non-technology methods rather than advanced PETs?",  
``are users concerned enough and aware of PETs to use them?" ,
``how were their previous experience with PETs?",
``how can we encourage users to adopt more advanced and dedicated PETs?"

\begin{RedundantContent}
Privacy settings, clear information/history and anti-malware are also in the common topmost preferred methods.
It will be valuable to further understand users' perception of the effectiveness of these preferred methods, to investigate the human-computer interaction involved, as well as possibly find ways to offer technology alternatives. 

There is also a $90\%$ similarity in most preferred sharing methods between the two countries. 
[\dots \dots how do these compare with the most preferred privacy methods - UD and BI privacy settings \& clear info/history are common in both countries. \dots \dots]
\end{RedundantContent}


\begin{RedundantContent}
A higher number of US participants reported use compared to the UK with statistical significance for $17$ privacy methods, with the UK only reporting higher use for Paypal instead of online banking. [\dots to discuss why US participants claim to use more privacy methods? \dots]
However, overall the privacy methods differences between the two samples only range between $5.3\%$ to $13.6\%$ of participants.

For example, about $10\%$ more US ($43.2\%$) participants chose not to use Facebook as privacy method compared to the UK (32.7\%), indicating that Americans may have responded slightly more to the abandon Facebook campaign (`\#DeleteFacebook'), following the Cambridge Analytica scandal~\cite{https://www.nytimes.com/2018/03/21/technology/users-abandon-facebook.html} in $2018$.
However, in both countries participants use this approach as privacy method, for example, the UK reports a drop of $30\%$ Facebook users in summer $2019$~\cite{https://www.telegraph.co.uk/technology/2019/07/06/exclusive-britons-abandon-facebook-usage-plummets-third/}. 

There is a statistical difference in preference for $12$ sharing method between the two countries, where the difference range between $5.7\%$ to $15.2\%$ participants, except for Reddit with $27.7\%$ more US participants and WhatsApp with $55.5\%$ more UK participants.
\end{RedundantContent}

\section{Conclusion}
This paper provides an initial investigation of a mapping between perceived privacy dis-empowerment online and preferences for privacy and sharing methods, as well as offers a cross-national investigation.
We identify a few non-technology privacy methods that are preferred over more advanced and standalone privacy-enhancing technologies. 
This raises questions for future research, in particular why individuals prefer methods that seem more accessible and integrated within non-privacy focused environments and non-technology methods, rather than more advanced and more technical privacy technologies.

%
%
%
 \bibliographystyle{splncs04}
\bibliography{repository,cfs,laser,psychology_affect,passwords,methods_resources,fearappeals,emotion,anger,mturk,empowerment,se,privacy,surveys,interventions,psychometry}
%

\section{Appendix}
\begin{table*}[h]
\centering
\caption{Privacy Method Differences across Countries with Chi Square Test, sorted by effect size $V$} 
\label{tab:differences}
\footnotesize
\resizebox{\textwidth}{!}{
\begin{tabular}{lllrrrlrrrrl@{}ll} 
\toprule
\multirow{2}{*}{\textbf{Privacy Method}} &\multirow{2}{*}{\textbf{CAT}} &\multirow{2}{*}{\textbf{\% Participants}}& \multicolumn{3}{c}{\textbf{\% in Country}} & &\multicolumn{3}{c}{\textbf{\% in Method}} & \multirow{2}{*}{\textbf{$X^2(2)$}} & \multirow{2}{*}{$p$} &&  \multirow{2}{*}{\textbf{Cramer $V$}}\\
\cline{4-6}
\cline{8-10}
&&& \textbf{US} & \textbf{UK} &\textbf{DE}&  & \textbf{US} & \textbf{UK} &\textbf{DE}& \\
\midrule
1 Pseudonyms &ANO &53.1 & 47.2&36.6&76.1&&29.6&23.0&47.4&101.087&.000&***&.334 \\
2 Anonymous Profiles &ANO &59.2& 57.8&44.2&75.7&& 32.6&25.0&42.5 &65.522&.000&***&.263\\
3 Have several emails&ANO&72.5&70.6&60.4&86.7&&32.5&27.8&39.7&53.343&.000&***& .243\\
4 NoScript &BHP&15.9&12.9&6.9 &27.9  &&   27.1&14.6&58.3&52.823&.000&***&.241\\
5 Give fake info &ANO&45.3& 42.2&33.7&60.1&&31.1&24.8&44.0&44.234&.000&***&.221 \\
6 Adblock &COP&76.5&77.6&65.0&87.0&&33.9&28.4&37.8&41.045&.000&***&.213\\
7 VPN&COP & 37.7& 32.7& 28.4&52.2&&28.9&25.1&45.9&41.250&.000&***&.213\\    
8 HTTPS &COP&68.1& 69.0&56.8 &78.7 &&  33.8&27.8&38.3&33.724&.000&***&.193\\
9 TOR &ANO & 13.5&10.6&7.6&22.3 && 26.2&18.9&54.9&31.172&.000&***&.185    \\
10 Virtual Machines & ANO &13.8&12.2&6.9&22.3&&29.6&16.8&53.6&30.803&.000&***&.184  \\
11 Anti-tracking extension &BHP&31.2&  30.7&21.1&41.9&&32.9&22.6&44.5&30.308&.000&***&.183\\
12 Not use Facebook &ANO&43.0& 43.2&32.7&53.2 && 33.6&25.4 &41.0&25.857&.000&***&.169 \\ 
13 Paypal instead of online banking & PLS & 74.6&66.3&74.6&83.1&&29.7&33.4&36.9&22.302&.000&***&.157 \\
14 Proxy & ANO & 26.7&22.1&21.5&36.5&&27.7&26.9&45.5 &22.438&.000&***&.157\\
15 Read terms of service & PLS & 44.2 &50.8&48.5&33.2&&38.4&36.7&24.9&22.385&.000&***&.157\\
16 Not access accts in public place & ANO & 66.7&73.9&69.3&56.8&&37.0&34.7&28.3&21.308&.000&***&.153\\
17 Request data collected & PLS& 19.1&17.8&12.5&26.9&&31.2&22.0&46.8&20.660&.000&***&.151\\
18 DuckDuckGo &BHP&21.5& 26.1&12.9&25.6&& 40.5&20.0&39.5&20.092&.000&***&.149\\
19 Ghostery & BHP & 11.8 & 10.2& 6.9 & 18.3 &&29.0&19.6&51.4&19.739&.000&***&.148\\
20 Kapersky & PLS& 14.1&9.9&11.6&20.9&&23.4&27.3&49.2&17.617&.000&***&.139\\
21 Firewall & COP& 74.9& 69.6&71.9&83.1&&31.1&32.1&36.8&16.504&.000&***&.135\\
22 Switch off camera &ANO&34.7& 68.0&56.8&69.4&&35.1&29.3&35.6&12.743&.002&**&.119 \\
23 Anti-spyware &PLS& 65.3&72.9&62.7&60.1&&37.3&32.1&30.6&12.241&.002&**&.116 \\
\bottomrule
\end{tabular}
}
These differences are statistically significant under Bonferroni correction.\\
Effect size \emph{Cramer} $V<.20$ corresponds to a weak effect, $.20<V<.40$ corresponds to a moderate effect.
\end{table*}

\end{document}